# REGULATING AI: APPLYING INSIGHTS FROM BEHAVIOURAL ECONOMICS AND PSYCHOLOGY TO THE APPLICATION OF ARTICLE 5 OF THE EU AI ACT *

**Huixin Zhong, Eamonn O'Neill, Janina A. Hoffmann**
Centre for Doctoral Training in Accountable, Responsible and Transparent AI, University of Bath, United Kingdom
{`hz877@bath.ac.uk, E.ONeill@bath.ac.uk, jah253@bath.ac.uk`}

## ABSTRACT

Article 5 of the European Union's Artificial Intelligence Act is intended to regulate AI use to prevent potentially harmful consequences. Nevertheless, applying this legislation practically is likely to be challenging because of ambiguously used terminologies and because it fails to specify which manipulation techniques may be invoked by AI, potentially leading to significant harm. This paper aims to bridge this gap by defining key terms and demonstrating how AI may invoke these techniques, drawing from insights in psychology and behavioural economics. First, this paper provides definitions of the terms "subliminal techniques", "manipulative techniques" and "deceptive techniques". Secondly, we identified from the literature in cognitive psychology and behavioural economics three subliminal and five manipulative techniques and exemplify how AI might implement these techniques to manipulate users in real-world case scenarios. These illustrations may serve as a practical guide for stakeholders to detect cases of AI manipulation and consequently devise preventive measures. Article 5 has also been criticised for offering inadequate protection. We critically assess the protection offered by Article 5, proposing specific revisions to paragraph 1, points (a) and (b) of Article 5 to increase its protective effectiveness.

## 1 Introduction

The European Commission first drafted the EU AI Act in April 2021. In June 2023, the members of the European Parliament solidified the negotiating position for the AI Act, marking a notable step forward in the development of the world's first legal framework for AI. Negotiations will commence with EU member countries to shape the final legislation. The goal is to reach a final agreement by the end of 2023 [1]. The EU AI Act proposes to apply a different regulatory standards to AI technologies depending upon the risk level of the technology rather than the type of technology itself. This risk based approach to AI applications aims to ensure the safety and trustworthiness of AI while encouraging innovation and development. The Act broadly categorizes AI technologies into four different risk levels. Unacceptable risk: This category includes AI technologies that pose a clear threat to an individual's safety, livelihood or rights, such as social ranking systems and manipulation of individuals. These types of AI technologies are prohibited. High Risk: This category includes AI technologies applied in critical infrastructures such as transport, education, employment and worker management. High risk AI technologies must meet strict standards before being deployed in the market. High-risk AI technologies are required to conduct adequate risk assessments and include mitigation systems to ensure the high quality and robustness of the products. Limited Risk: This category refers to AI technologies that must meet specific transparency obligations when interacting with users, such as chatbots. Minimal or No Risk: This category includes AI technologies that pose minimal risks to consumers, such as AI-powered video games.

The EU AI Act is structured in Articles, each addressing the regulation of AI technologies based on their corresponding risk levels. Article 5 of the Act is dedicated to the regulation of AI technologies that could present unacceptable risks, thus justifying their prohibition in the EU market. In particular, Article 5, paragraph 1, points (a) and (b) are focused

---

on the regulation of AI manipulation that could affect the general public and vulnerable groups. These terms were originally expressed as follows [2]:

> (a) the placing on the market, putting into service or use of an AI system that deploys subliminal techniques beyond a person's consciousness in order to materially distort a person's behaviour in a manner that causes or is likely to cause that person or another person physical or psychological harm;
>
> (b) the placing on the market, putting into service or use of an AI system that exploits any of the vulnerabilities of a specific group of persons due to their age, physical or mental disability, in order to materially distort the behaviour of a person pertaining to that group in a manner that causes or is likely to cause that person or another person physical or psychological harm.

However, Article 5 has been criticized for lack of clarity [3], inadequate protective measures [4], and failure to achieve its stated objectives [5]. Partly in response to these criticisms, the EU Parliament passed amendments to the original draft EU AI Act on 14 June, 2023 [6]. In the amendments, the EU Parliament made the following revisions to Article 5, paragraph 1, points (a) and (b):

> (a) the placing on the market, putting into service or use of an AI system that deploys subliminal techniques beyond a person's consciousness or purposefully manipulative or deceptive techniques, with the objective to or the effect of materially distorting a person's or a group of persons' behaviour by appreciably impairing the person's ability to make an informed decision, thereby causing the person to take a decision that that person would not have otherwise taken in a manner that causes or is likely to cause that person, another person or group of persons significant harm. The prohibition of AI system [sic] that deploys subliminal techniques referred to in the first sub-paragraph shall not apply to AI systems intended to be used for approved therapeutical purposes on the basis of specific informed consent of the individuals that are exposed to them or, where applicable, of their legal guardian.
>
> (b) the placing on the market, putting into service or use of an AI system that exploits any of the vulnerabilities of a person or a specific group of persons, including characteristics of such person's or a [sic] such group's known or predicted personality traits or social or economic situation age, physical or mental ability with the objective or to the effect of materially distorting the behaviour of that person or a person pertaining to that group in a manner that causes or is likely to cause that person or another person significant harm.

This amendment acknowledges in point (a) the insufficiency of protective measures due to the narrow interpretation of 'subliminal techniques'. Consequently, it broadens the scope to encompass both 'manipulative techniques' and 'deceptive techniques'. Nevertheless, the amendment does not explicitly define the terminologies employed in this Article, which may hinder the practical implementation of the Act. In addition to clearer definitions, stakeholders might benefit from an illustration of these techniques via case studies to detect and prevent AI-based manipulation of users. When the initial draft was published, researchers such as [5, 3] attempted to clarify the terminologies used in Article 5, but rarely provided illustrative examples of these techniques. To address these gaps, this paper employs insights from behavioural economics and psychology to enhance the clarity and protective capacity of Article 5. We exemplify subliminal and manipulative techniques in case studies. Further, we point towards instances in which the provision of Article 5 may remain inadequate to safeguard the general public and vulnerable groups. Our research aims to support the ongoing drafting of the EU AI Act theoretically, assist in its refinement, and prepare for the upcoming legal implementation.

The paper is structured as follows. In Section 2, we clarify the ambiguous terminology in Article 5 following insights from behavioural economics and psychology and describe case studies applying subliminal and manipulative techniques. In Section 3, we identify limitations of Article 5 in providing protection to both the general public and vulnerable groups. We propose specific revisions to Article 5, paragraph 1, points (a) and (b).

## 2 Terminology in Article 5: Insights from Behavioural Economics and Psychology

In the most recent amendments to point (a), three terms may hold considerable importance in translating point (a) from legal provision to practice: subliminal techniques, manipulative techniques and deceptive techniques. Interpreting these terms correctly is key to the successful application of the EU AI Act in practice and to achieving the protective capacity of the Act. In this section, we interpret these key terms from the perspective of psychology and behavioural economics. We illustrate subliminal and manipulative techniques with historical examples, discuss their impact on individual behavior, and outline potential scenarios where AI systems could invoke these techniques to manipulate humans.

## 2.1 Subliminal Techniques

In the field of cognitive psychology, subliminal techniques refer to methods that aim to influence individuals using stimuli below an individual's perception thresholds [7]. Subliminal techniques gained popularity in the 1950s, when initial studies suggested that subliminally presented words or sounds in advertisements could influence consumer behaviour [8]. However, these studies suffered from design flaws and inappropriate measures of unawareness [9, 10]. While some research has subsequently demonstrated the effectiveness of subliminal techniques in practical applications in education [11], healthcare [12] or advertising [13], the idea that behavior can be influenced unconsciously remains controversial[14, 15].

In this paper, we adopted a modified RRMG rapid review methodology [16] to quickly collate and synthesize findings on subliminal techniques while adhering to methodological rigor. We searched for articles in Google Scholar, ACM Digital Library, and IEEE Xplore from 2013 to June 2023. We reviewed title and abstract of 51 entries, among which we identified three systematic reviews on subliminal techniques [17, 18, 7]. Across these review articles, three subliminal techniques appeared most frequently and we then searched for analogous AI applications. The forthcoming sections of the paper present these three subliminal techniques and provide examples of how these techniques have been exploited previously to influence people's decision making and behaviour. Lastly, we discuss cases in which AI might exploit these subliminal techniques.

**Tachistoscopic presentation**

In tachistoscopic presentation, visual stimuli are displayed for extremely short duration with the goal to unconsciously influence attitudes and behavior [19]. In 1957, Vicary first applied tachistoscopic presentation in a movie theater with 45,699 participants [19]. A tachistoscopic device was used to flash the phrases "Drink Coca-Cola" and "Eat popcorn" for 1/3000th of a second every 5 seconds throughout a film. The results showed an 18.1% rise in Coca-Cola sales and a 57.7% increase in popcorn sales [13]. However, the validity of Vicary's experiment has been a point of contention, and its scientific integrity has been frequently challenged.

Despite the controversies, in recent years researchers have begun to explore if tachistoscopic presentation can benefit educational outcomes and enhance individual learning. For instance, [20] integrated tachistoscopic presentation into a computer-based tutoring system to teach users the tricks of an old magic box. Participants in the experimental group were briefly shown a 33.33 ms flash of the correct answer between two masked slides. Findings revealed that this group (who saw the correct subliminal answer) outperformed both the control group (who received no subliminal answer) and the miscue group (who were shown subliminal indications of incorrect answers). It is worth noting that the duration and the content of subliminal stimuli was still intentionally chosen by humans, not by AI. However, it is possible that AI can be used to determine the optimal presentation duration and stimulus content so that tachistoscopic presentation steers an individual's decisions more effectively. We will elaborate on this point after introducing two other subliminal techniques.

**Masked Stimulus**

The masked stimulus technique can be employed through both visual and auditory channels. Masking diminishes the intensity of a stimulus or alters its perception, for instance by presenting multiple stimuli simultaneously. Visual masking can encompass brightness, texture, frequency, time and colour masking [21]. Auditory masking can involve speed and backward message masking [21]. In speed masking, the original recording is accelerated to such an extent that the content cannot be consciously processed. Conversely, in backward masking, the primary message is played in reverse, often hidden behind another one. For instance, a message might be hidden underneath a song and only discernible when the song is played in reverse [22].

**Conceptual Priming**

In conceptual priming, individuals are exposed to stimuli conveying a certain meaning. These primes activate associated memories and, in turn, are thought to influence subsequent actions. Famous examples of conceptual priming span from exposure to money, to priming the concept of intelligence to prinming stereotypes of old age. For instance, Bargh[23] exposed one group of participants to a set of words associated with old age, such as "wrinkles, bitter and alone". On average, these participants were found to walk more slowly to the elevator than a group of control participants. Yet, attempts to replicate Bargh's results and other key findings from the literature have been unsuccessful [24, 25].

As a result, it remains an open question to what degree conceptual priming is an effective technique to influence an individual's actions and decisions [26, 27].

**The AI Advantage**

To date, there is no concrete evidence that AI has already exploited subliminal techniques to manipulate humans. Following the conflicting findings on subliminal techniques and the difficulties in achieving effective stimulation thresholds, some researchers suggested to remove subliminal techniques from Article 5 [4, 3]. Still, AI might possess the potential to elevate the effectiveness of subliminal techniques.

To successfully exploit subliminal techniques, two factors are of great importance: identifying the most effective perception threshold, such as the duration and frequency of the presentation, and determining the most effective hidden content. It is well known that perceptual thresholds as well as the most effective cues vary between individuals. AI may be able to customise subliminal cues through "micro-targeting" and through continuously collecting vast amounts of individuals' information including perception, behaviour, personality and preferences across multiple social media platforms. AI may also be able to fine-tune presentation duration to perception thresholds of individual users, spread a variety of subliminal cues and collect people's feedback to find the most effective cues. Finally, AI can continuously refine user profiles and deliver precise content 24 hours a day among a large number of people, achieving a repeated targeted manipulation of user behaviour. In conclusion, the key advantages AI has in influencing people through subliminal techniques are micro-targeting, large-scale application, and the relentless collection of feedback and dissemination of tailored information [5], possibly rendering the use of subliminal techniques more effective on a large scale.

## 2.2 Manipulative and Deceptive Techniques

While subliminal techniques have faced doubts regarding their effectiveness, AI may potentially manipulate people by exploiting manipulative or deceptive techniques. Cohen differentiated deception from manipulation: "while the latter compromises good judgment by interfering with its standard functioning and by inducing suboptimal judgment (interfering with its 'form'), deceptions undermine judgment by interfering with its input ('content'): they do not change its standard functioning; rather, they prevent its successful conclusion by feeding it the wrong data. " [28, p. 486]. Thus, we define manipulation as distorting the form or structure of the judgment process, leading to outcomes that may not be in the best interests of the decision maker. We define deception as producing false information to distort the 'content' of decision making, leading to outcomes that may not be in the best interests of the decision maker. Several deception techniques have been distinguished in previous reviews into the categories "imitating, obfuscating, tricking, calculating and reframing" [29]. The potential of AI systems to apply deceptive techniques has been extensively outlined in reviews [29] and its implications have been delineated thoroughly [30, 31]. Given the wealth of existing research on deception, here we primarily concentrate on understanding and analysing manipulative techniques employed by AI.

In behavioral economics, the idea that subtle changes in the structure of a decision problem can cause a substantial shift in the decision process and, ultimately, individuals' choices has been coined as "nudging" [32]. Nudging aims to enable better decisions by triggering an individual's intuitive decision processes, so called heuristics and biases, via changes in the decision context [33]. Nudging shares with manipulative techniques the concept that the decision architect can successfully interfere with or distort the form of decision making. In the last decades, nudging has been extensively employed in economics to guide citizens towards making better decisions [32].Although the large-scale impact of nudging has been recently heatedly debated [34, 35, 36], the insights from nudging research may inform our understanding about which human mental shortcuts (or heuristics) AI might invoke to achieve manipulation.

The heuristics-and-biases approach, pioneered by Teversky and Kahneman [37], distinguishes between two modes or systems of thinking [38, 39]. In dual-process theory, System 1, also known as the heuristic-thinking system, operates swiftly and effortlessly, often without conscious awareness. Although this system is generally adequate for decision making, it can sometimes compromise accuracy in favour of speed and efficiency. In contrast, System 2 is thought to engage in complex problem solving that requires attention and conscious deliberation. Decisions following System 2 are supposed to take longer, but may be more accurate.

Several heuristics have been proposed to underpin heuristic thinking. For stakeholders pertinent to the EU AI Act, such as AI developers, algorithm auditors and legal practitioners, it is of paramount importance to understand the manifestations of these heuristics, rather than merely acquiring an understanding of abstract terminologies. Consequently, we present five classical heuristics identified through rigorous experimentation in psychology. Nudging approaches in behavioural economics have often aimed to elicit these heuristics to facilitate behavioural change. It is thus possible that AI similarly invokes those heuristics when applying manipulative techniques. We provide an explanation of each heuristic, supplemented with examples, to illustrate how AI can use manipulative techniques to alter individuals' behaviour.

**Representativeness Heuristic**

The representativeness heuristic proposes that individuals estimate the likelihood of an event based on its similarity to an existing stereotype or model [37]. People may follow the representativeness heuristics to replace a more complex Bayesian calculation. For instance, an individual might infer that a person who is described as introverted and enjoys reading is more likely to be a librarian than a salesperson, although the population of salespeople significantly outnumbers librarians. This inference is drawn because the description of an introverted person aligns more closely with the stereotypical image of a librarian than that of a salesperson [40].

In social media research, the representativeness heuristic has been suggested as one mechanism through which echo chambers may be generated, thereby causing behavior change[41]. First, if individuals follow the representativeness heuristic, this increases the likelihood that individuals will adopt and spread information that matches their pre-existing beliefs [42]. Second, it prompts people to share their views primarily with those who are already in agreement, leading to a reinforcement of similar perspectives and the exclusion of opposing ones [43].

On social media and recommendation platforms, machine learning and AI systems may be implemented to provide personalised recommendations and filtered information. The initial aim and the positive side of designing personalised and filtered recommendations is to enhance users' engagement [44] and to reduce information overload [45]. However, the downside of such tailored content is that it can reinforce individuals' existing biases and stereotypes and influence individuals' decisions through representativeness heuristics. AI systems on social media and recommendation platforms may invoke the representativeness heuristic to create echo chambers in order to shape individuals' political beliefs. These echo chambers can amplify polarisation and radicalism [44]. For instance, if an AI system detects a user with slight racist inclinations based on social media data, it might consistently recommend more extreme racist content to that user. Over time, the user might come to see such views as normal and widespread within their social group, mistakenly believing that most of their peers share these views. The potential dangers of AI-driven social media and recommendation platforms in terms of increasing polarisation and racism have been extensively discussed. Haroon and colleagues [44] conducted a large-scale study which provided empirical evidence of ideological bias in YouTube's recommendations, with even more pronounced radicalisation among right-leaning users.

**Availability Heuristic**

The availability heuristic proposes that individuals assess a specific topic, concept, method or decision, based on immediately accessible examples that come to mind. This heuristic operates on the premise that if information can be readily recalled, it must hold significance, or at least be more important than alternatives that are not so easily remembered. Consequently, the availability heuristic may lead individuals to overweigh the importance of recent information, resulting in a bias towards the most current news [37]. For example, when asked to estimate the probability of various causes of death, people often overestimate the probability of events that are frequently reported in the news, such as terrorist attacks or plane crashes, because these events are more readily recalled. Conversely, they underestimate the likelihood of more common but less reported causes of death, such as heart disease or car accidents [46]. Thaler and Sunstein [32] argue that the availability heuristic can be invoked to nudge the public or individuals towards taking precautions against potential risks. For example, during the COVID-19 pandemic, the frequent dissemination of stories about severely affected individuals, coupled with images of overcrowded hospitals and healthcare workers in protective gear, made these scenarios readily available in people's minds. This heightened perception of risk may have encouraged precautionary behaviours such as mask wearing, social distancing and vaccination [47].

AI can make this process more influential by targeting these narratives to individuals who are less likely to take precautions. For example, a social media algorithm could prioritize the precautionary content of COVID-19 in the feeds of users whom the algorithm identified as less likely to adhere to safety measures based on their online behaviour and preferences. On the dark side, industry may also employ AI to nudge consumer behaviour. In 2022, [48] investigated how 12 leading food and beverage companies use AI to increase sales of unhealthy food. The research revealed that these companies are using AI to track real time sales data in stores, to adjust the stock of products and to automatically replenish popular items. As a result, consumers frequently encounter these popular products, making them easily accessible. This consistent exposure might lead consumers to overestimate the popularity of certain products, influencing their purchasing decisions. While research shows that implementing such AI systems leads to greater consumption of those products and boost brand loyalty, it also suggests a decrease in consumers' inclination to buy healthy foods [48].

**Anchoring Effect**

The anchoring effect describes the tendency of individuals to rely overly on the initial piece of information they encounter, known as the "anchor". Once an anchor is established, subsequent judgments are made by adjusting away from this anchor, resulting in a biased interpretation of other information in relation to the anchor [37]. For instance, during a house price negotiation, if the seller sets the initial price (the anchor), subsequent negotiations are likely to revolve around this initial figure, even if it is significantly above the market value. As a result, the initial price can possess a profound influence on the negotiation process.

AI can invoke the anchoring effect to influence human decision making in various contexts. In a commercial environment, AI can utilize data on an individual's previous purchases or browsing history to establish personalized price anchors. For example, if an AI system identifies that a user has a history of purchasing high-end products, it might set a higher anchor price for similar products in the future. In the context of news recommendation systems, AI algorithms can employ the anchoring effect to shape people's perceptions. By presenting a particular viewpoint or piece of information first (the anchor), the AI can influence how users interpret subsequent information.

**Status Quo Bias**

Status quo bias describes the tendency of individuals to prefer the existing state of affairs. The current situation, or status quo, is perceived as a reference point, with any deviation from this point viewed as a loss [49]. This bias can result in scenarios where individuals maintain the status quo in their decision-making, even when change could be advantageous. The status quo bias manifests in various contexts, including financial decisions, health choices, and public policy preferences. For instance, individuals may retain their existing health insurance plan, even if superior options are available, due to their inclination towards the status quo [32]. Similarly, investors may retain underperforming assets due to their reluctance to alter their current portfolio.

Some AI systems may similarly invoke the status quo bias to influence compliance. For example, Uber's AI system leveraged status quo bias to subtly encourage drivers to work longer hours. As drivers near the end of their shift and attempt to log out of the Uber system, the AI system sends personalized notifications about high demand in their current area, often accompanied by increasing surge pricing icons. Furthermore, before one ride concludes, the AI automatically queues up the next ride. This setup leads drivers to continue without taking breaks, as they are inclined to maintain the status quo, which in this case is the seamless continuation of rides arranged by the system [50].

**Social Conformity**

Social conformity refers to the phenomenon that individuals adjust their judgments and behaviors to align with those of a group, either to enhance the accuracy of their decisions or to gain acceptance within the group [51]. Across a variety of domains, nudging research has sought to change behavior by providing social information that, in turn, elicits social conformity. For instance, in online shopping platforms, products that have received higher ratings from consumers are more likely to be perceived as high quality and chosen by subsequent consumers.

AI technologies can invoke social conformity at both the content level and the human-AI interaction level. At the content level, machine learning techniques can highlight and recommend products that are popular and have received the most clicks, thereby guiding consumer behavior. From the perspective of human-AI interaction, recent studies suggest that individuals are more likely to accept advice from multiple AI systems rather than a single AI [52, 53]. On the flipside, it is possible that multiple AI systems that disseminate consistent fake news could be perceived as even more trustworthy than a single AI system. Recent advances in large language models, such as OpenAI's Chat-GPT and Google's PaLM, allow individuals nowadays to seek out opinions and advice from multiple AI systems. However, it is still an open question how users will in future respond to and interact with multiple AI systems.

### 2.2.1 Nudging Versus Manipulation

Nudging citizens to adopt heuristics that ultimately encourage individuals to make better choices is a widely recognized practice in behavioral economics [32]. The study of how AI technologies can harness different heuristics to aid and boost human decisions is still in its infancy, necessitating comprehensive human-AI interaction experiments, but also an ethical debate within society about its social acceptability. Insights from nudging research may inform this debate and help identify which aspects of AI systems need to be communicated to the general public to render them more transparent and acceptable. Although the ethical debate surrounding the notion that 'nudging' impinges upon an individual's autonomy of choice remains unresolved, some consensus has been reached to differentiate between nudging and manipulation. The first criterion for ethical nudging is related to its social acceptability. The degree to which nudging is socially acceptable often depends on the specific context, cultural and social roots, which complicates the establishment of a universal standard [54]. Moreover, the question of who should have the right to nudge the general public also sparks a serious debate. A less controversial and more practical requirement for ethical nudging may be the 'transparency requirement.' Nudging recipients must be transparently informed and provide their consent. Translating these criteria to the field of AI implies that if AI technologies aim to invoke certain heuristics or apply subliminal techniques, this practice should be transparently communicated to the users and user consent should be obtained.

# 3 Enhancing Protection for the General Public and Vulnerable Groups

## 3.1 Enhancing Protection of the General Public

Beyond the use of ambiguous terminology, the phrasing of the recent amendments of the EU AI Act may still lead to an insufficient level of protection for the general public. Although the wording of Article 5, paragraph 1, point (a) has changed from the previous draft's "deploys subliminal techniques ... in order to materially distort a person's behaviour" to the revision's "deploys subliminal techniques ... or purposefully manipulative or deceptive techniques, with the objective to or the effect of materially distorting a person's ... behaviour", there is still a focus on subjective intention. Manipulative techniques do not necessarily need to be deliberately deployed, and can also be invoked without intention. The current use of "deploy" in the EU AI Act may exempt scenarios where AI developers assert that they did not intentionally deploy techniques to manipulate people but merely unintentionally invoked people's heuristics. This could potentially diminish the protective efficacy of Article 5, paragraph 1, point (a) for the general population. Therefore, we suggest replacing "purposefully" by "invokes" before "manipulative or deceptive techniques".

Thus, the revised text of Article 5, paragraph 1, point (a) would include: "The placing on the market, putting into service or use of an AI system that deploys subliminal techniques beyond a person's consciousness or invokes manipulative or deceptive techniques, with the objective to or the effect of materially distorting a person's or a group of persons' behaviour by appreciably impairing the person's ability to make an informed decision, thereby causing the person to take a decision that that person would not have otherwise taken in a manner that causes or is likely to cause that person, another person or group of persons significant harm". We argue that the Act should also explicitly mandate AI developers to undertake comprehensive AI audits and human-AI interaction reviews prior to the market introduction of products, in order to assess the presence and effects of manipulative techniques, and should mandate product manufacturers to monitor market and consumer reactions after the products are launched.

## 3.2 Enhancing Protection of Vulnerable Groups

Just as the focus on the subjective intention to cause harm diminishes the protective capacity of Article 5, paragraph 1, point (a), point (b) may also offers insufficient protection of vulnerable groups due to a similar emphasis on subjective intention. AI systems do not need to exploit vulnerabilities to cause significant harm to vulnerable groups. Systems that merely overlook the vulnerabilities of these groups could potentially cause significant harm. For instance, individuals with Autism Spectrum Disorders (ASDs) often struggle with understanding non-literal speech, such as irony or metaphor, due to impairments in social understanding and recognizing the speaker's communicative intention [55, 56]. In recent years, chatbots have become popular that engage with and train individuals with ASDs to enhance their social skills. If a chatbot is trained solely on a database of typical adult conversations, it may incorporate elements such as jokes and metaphors that individuals with ASDs may interpret literally and act upon, potentially leading to significant harm.

Moreover, chatbots have been widely introduced mental health applications [57]. The goal is to provide psychological therapy to individuals with mental health disorders. As large-language models such as ChatGPT continue to advance at a rapid pace, numerous companies are exploring their integration into mental health chatbots. However, such applications may risk doing more harm than good if they fail to account for the unique vulnerabilities of mental health patients. For example, one of the most crucial principles in administering psychological therapy is consistency [58]. This notion encompasses not only the continuity of consultation sessions but also the preservation of a consistent treatment pattern. This pattern demands an explicit, coherent framework to guide the entirety of the psychological therapy process. The consistency principle is particularly significant for patients with mental health conditions such as Borderline Personality Disorder (BPD) [59]. BPD patients typically fear inconsistency in communication patterns, which may trigger feelings of insecurity and abandonment [59]. However, large language models struggle to maintain a unique and consistent identity in their responses [57]. The responses of large language models often appears inconsistent, as the model relies on a wide array of pre-existing data to generate replies. Such inconsistent responses may amplify BPD patients symptoms and potentially cause significant harm.

Furthermore, patients suffering from mental health disorders such as depression, who may have suicidal beliefs, may require timely professional intervention. However, research suggests that the use of mental health chatbots may cause patients to avoid real life interactions and rely solely on these chatbots [57]. Such behaviour may result in delayed consultation with professional counsellors and deferred treatment, potentially contributing to escalation to severe consequences such as suicide.

Therefore, we propose further amending Article 5, paragraph 1, point (b) beyond the current revision by the addition of the following: "AI systems designed for use by a vulnerable group should be tailored to the unique characteristics of that group. If an AI system specifically designed for such a group fails to meet the tailored requirements, its entry into the market should be prohibited".

# 4 Conclusion

In this paper, we clarified and interpreted the ambiguous terminologies presented in Article 5 of the EU AI Act. Subliminal techniques can be understood as methods that aim to influence people by employing stimuli that are below the conscious perception threshold of the individual. Manipulative techniques can be defined as techniques that distort the form or structure of decision making, leading to outcomes that may not be in the best interests of the decision maker. Deceptive techniques may be defined as techniques that produce false information to distort the content of decision making, leading to outcomes that may not be in the best interests of the decision maker.

Furthermore, we have provided examples of three common subliminal techniques and five classical heuristics that may be invoked by AI to alter people's behaviours. We identified tachistoscopic presentation, masked stimulus, and conceptual priming as the most influential subliminal techniques. Further, we pointed out that manipulative techniques applied by AI systems might invoke similar heuristics and cognitive phenomena as behavioral economists intend to affect via nudging. These may include representativeness heuristic, availability heuristic, anchoring effect, status quo bias, and social conformity. This list is intended to serve as a practical guide for stakeholders such as AI developers, algorithm auditors, users and legal practitioners, enabling them to recognise these techniques and identify appropriate countermeasures. However, it is important to note that this list is not meant to be exhaustive. Future research may expand the catalogue of subliminal, manipulative and deceptive techniques that AI system exploit.

Finally, we believe that the current provisions of the EU AI Act do not provide sufficient protections because they place too much emphasis on subjective intentions and fail to address situations where techniques are invoked by AI without the specific intention of the AI developer. Therefore, we propose further amendments to the Act to enhance its protective efficacy. In our proposed revisions, applications that cause similar harmful consequences as intentional misconduct would also be classified as posing unacceptable risks, even if these risks are not subjectively intended. This inclusion would expand the range of risks deemed unacceptable and strengthen the protective measures under Article 5, compelling AI product developers to undertake comprehensive testing before market release. However, it is important to remember that one of the primary goals of the AI Act is to balance regulation and innovation. So, while stakeholders may take our recommendations into account, they should also reflect further on how best to strike this delicate balance.

Overall, this paper aims to help bridge the gap between the legal provision and its practical application by applying insights from behavioural economics and psychology. Our work contributes to the ongoing discourse on AI regulation, providing a practical guide for interpreting, applying and improving Article 5 of the EU AI Act.

# 5 Acknowledgments

This work was supported by the UKRI Engineering and Physical Sciences Research Council [grant number: 2284306] and the University of Bath. We would like to express our gratitude to Dr. Charles Larkin from the Institute for Policy Research (IPR) at the University of Bath, Sifan Yu from the Law School at Bristol University, and Syeda Zahra from the Centre for Doctoral Training in Accountable, Responsible, and Transparent AI at the University of Bath, for their valuable and constructive feedback on the initial draft of our paper.

# References

[1] European Parliament. Eu ai act: First regulation on artificial intelligence. https://www.europarl.europa.eu/news/en/headlines/society/20230601ST093804/eu-ai-act-first-regulation-on-artificial-intelligence, 2023. Accessed: 2023-08-03.

[2] European Commission. The eu artificial intelligence act. https://artificialintelligenceact.eu/the-act/, 2023. Accessed: 2023-05-19.

[3] Matija Franklin, Hal Ashton, Rebecca Gorman, and Stuart Armstrong. Missing mechanisms of manipulation in the eu ai act. In *The International FLAIRS Conference Proceedings*, volume 35, 2022.

[4] R Uuk. Manipulation and the ai act. *The Future of Life Institute*, 2022.

[5] Tegan Cohen. Regulating manipulative artificial intelligence. *SCRIPTed*, 20:203, 2023.

[6] European Parliament. Amendments of eu ai act. https://www.europarl.europa.eu/doceo/document/TA-9-2023-0236_EN.html, 2023. Accessed: 2023-08-03.

[7] Athanasios Drigas, Eleni Mitsea, and Charalampos Skianis. Subliminal training techniques for cognitive, emotional and behavioral balance. the role of emerging technologies. *Technium Soc. Sci. J.*, 33:164, 2022.

[8] James V McConnell, Richard L Cutler, and Elton B McNeil. Subliminal stimulation: An overview. *American psychologist*, 13(5):229, 1958.

[9] Charles W Eriksen. Discrimination and learning without awareness: a methodological survey and evaluation. *Psychological review*, 67(5):279, 1960.

[10] William Bevan. Subliminal stimulation: A pervasive problem for psychology. *Psychological Bulletin*, 61(2):81, 1964.

[11] LH Silverman and R Grabowski. The effects of activating oneness fantasies on the anxiety level of male and female college students. *Unpublished manuscript, Research Center for Mental Health, New York University*, 1982.

[12] Dennis P Saccuzzo and Donald L Schubert. Backward masking as a measure of slow processing in schizophrenia spectrum disorders. *Journal of Abnormal Psychology*, 90(4):305, 1981.

[13] Paul Messaris. How to make money from subliminal advertising and motivation research. *International Journal of Communication*, 7:16, 2013.

[14] Mary L Still and Jeremiah D Still. Subliminal techniques: considerations and recommendations for analyzing feasibility. *International Journal of Human–Computer Interaction*, 34(5):457–466, 2018.

[15] Michelle R Nelson. The hidden persuaders: Then and now. *Journal of Advertising*, 37(1):113–126, 2008.

[16] Chantelle Garritty, Gerald Gartlehner, Barbara Nussbaumer-Streit, Valerie J King, Candyce Hamel, Chris Kamel, Lisa Affengruber, and Adrienne Stevens. Cochrane rapid reviews methods group offers evidence-informed guidance to conduct rapid reviews. *Journal of clinical epidemiology*, 130:13–22, 2021.

[17] Nittaya Wongtada et al. Subliminal persuasion on a consumer's cognitive process: a review. *Journal of Applied Economic Sciences (JAES)*, 14(65):804–817, 2019.

[18] Angela Madan, Mihai Ioan Rosca, and Mirela Bucovicean. Theoretical approach of subliminal advertising. In *Eurasian Business and Economics Perspectives: Proceedings of the 31st Eurasia Business and Economics Society Conference*, pages 293–302. Springer, 2021.

[19] Timothy E Moore. The case against subliminal manipulation. *Psychology & Marketing*, 5(4):297–316, 1988.

[20] Pierre Chalfoun and Claude Frasson. Showing the positive influence of subliminal cues on learner's performance and intuition: an erp study. In *Intelligent Tutoring Systems: 10th International Conference, ITS 2010, Pittsburgh, PA, USA, June 14-18, 2010, Proceedings, Part II 10*, pages 288–290. Springer, 2010.

[21] J. Bu and B. Zheng. The innovation of ideological and political education in a perspective of subliminal message technology. In *Proceedings of the International Workshop on Artificial Intelligence and Education*, pages 13–17, November 2019.

[22] D. L. Rosen and S. N. Singh. An investigation of subliminal embed effect on multiple measures of advertising effectiveness. *Psychology and Marketing*, 9(2):157–173, 1992.

[23] John A Bargh, Mark Chen, and Lara Burrows. Automaticity of social behavior: Direct effects of trait construct and stereotype activation on action. *Journal of Personality and Social Psychology*, 71(2):230–244, 1996.

[24] Stéphane Doyen, Olivier Klein, Cora-Lise Pichon, and Axel Cleeremans. Behavioral priming: it's all in the mind, but whose mind? *PloS one*, 7(1):e29081, 2012.

[25] Tom Chivers. What's next for psychology's embattled field of social priming. *Nature*, 576(7786):200–202, dec 2019.

[26] Joseph Cesario. Priming, Replication, and the Hardest Science. *Perspectives on Psychological Science*, 9(1):40–48, jan 2014.

[27] Stéphane Doyen, Olivier Klein, Daniel J. Simons, and Axel Cleeremans. On the Other Side of the Mirror: Priming in Cognitive and Social Psychology. *Social Cognition*, 32(Supplement):12–32, jun 2014.

[28] Shlomo Cohen. Manipulation and deception. *Australasian Journal of Philosophy*, 96(3):483–497, 2018.

[29] Peta Masters, Wally Smith, Liz Sonenberg, and Michael Kirley. Characterising deception in ai: A survey. In *Deceptive AI: First International Workshop, DeceptECAI 2020, Santiago de Compostela, Spain, August 30, 2020 and Second International Workshop, DeceptAI 2021, Montreal, Canada, August 19, 2021, Proceedings 1*, pages 3–16. Springer, 2021.

[30] Eleftherios Chelioudakis. Deceptive ai machines on the battlefield: Do they challenge the rules of the law of armed conflict on military deception? *Available at SSRN 3158711*, 2017.

[31] Don Monroe. Deceiving ai. *Communications of the ACM*, 64(6):15–16, 2021.

[32] Richard H Thaler and Cass R Sunstein. *Nudge: The final edition*. Yale University Press, 2021.

[33] Luca Congiu and Ivan Moscati. A review of nudges: Definitions, justifications, effectiveness. *Journal of Economic Surveys*, 36(1):188–213, 2022.

[34] Nick Chater and George Loewenstein. The i-frame and the s-frame: How focusing on individual-level solutions has led behavioral public policy astray. *Behavioral and Brain Sciences*, 46:e147, 2023.

[35] Stephanie Mertens, Mario Herberz, Ulf J.J. Hahnel, and Tobias Brosch. The effectiveness of nudging: A meta-analysis of choice architecture interventions across behavioral domains. *Proceedings of the National Academy of Sciences of the United States of America*, 119(1):1–10, 2022.

[36] Maximilian Maier, František Bartoš, T. D. Stanley, David R. Shanks, Adam J.L. Harris, and Eric Jan Wagenmakers. No evidence for nudging after adjusting for publication bias. *Proceedings of the National Academy of Sciences of the United States of America*, 119(31):10–11, 2022.

[37] Amos Tversky and Daniel Kahneman. Judgment under uncertainty: Heuristics and biases: Biases in judgments reveal some heuristics of thinking under uncertainty. *science*, 185(4157):1124–1131, 1974.

[38] Daniel Kahneman. A Perspective on Judgment and Choice: Mapping Bounded Rationality. *American Psychologist*, 58(9):697–720, 2003.

[39] Daniel Kahneman and Shane Frederick. A model of heuristic judgment. In Keith J. Holyoak and Robert G. Morrison, editors, *The Cambridge Handbook of Thinking and Reasoning*, pages 267–293. Cambridge University Press, 2005.

[40] Daniel Kahneman. *Thinking, fast and slow*. macmillan, 2011.

[41] Giovanni Luca Ciampaglia, Alexios Mantzarlis, Gregory Maus, and Filippo Menczer. Research challenges of digital misinformation: Toward a trustworthy web. *AI Magazine*, 39(1):65–74, 2018.

[42] Christos Kyriacou and Nicos Stylianou. Probability fixed points,(in) adequate concept possession and covid-19 irrationalities. *Philosophical Psychology*, pages 1–25, 2023.

[43] Satrajit Ghosh Chowdhury. *Understanding Mis-and Dis-Information Consumption in a Polarized Society– Analyzing Selective Evaluation, Subjective Perception of Opinion Leaders and Effects of Heuristic Cues in Post-decision*. Ohio University, 2021.

[44] Muhammad Haroon, Anshuman Chhabra, Xin Liu, Prasant Mohapatra, Zubair Shafiq, and Magdalena Wojcieszak. Youtube, the great radicalizer? auditing and mitigating ideological biases in youtube recommendations. *arXiv preprint arXiv:2203.10666*, 2022.

[45] Eli Pariser. *The filter bubble: What the Internet is hiding from you*. penguin UK, 2011.

[46] Sarah Lichtenstein, Paul Slovic, Baruch Fischhoff, Mark Layman, and Barbara Combs. Judged frequency of lethal events. *Journal of experimental psychology: Human learning and memory*, 4(6):551, 1978.

[47] Caroline M Poland, Allison KS Matthews, and Gregory A Poland. Improving covid-19 vaccine acceptance: Including insights from human decision-making under conditions of uncertainty and human-centered design. *Vaccine*, 39(11):1547, 2021.

[48] Shane Joachim, Abdur Rahim Mohammad Forkan, Prem Prakash Jayaraman, Ahsan Morshed, and Nilmini Wickramasinghe. A nudge-inspired ai-driven health platform for self-management of diabetes. *Sensors*, 22(12):4620, 2022.

[49] Daniel Kahneman, Jack L Knetsch, and Richard H Thaler. Anomalies: The endowment effect, loss aversion, and status quo bias. *Journal of Economic perspectives*, 5(1):193–206, 1991.

[50] Daniel Susser, Beate Roessler, and Helen Nissenbaum. Online manipulation: Hidden influences in a digital world. *Geo. L. Tech. Rev.*, 4:1, 2019.

[51] Robert B Cialdini and Noah J Goldstein. Social influence: Compliance and conformity. *Annu. Rev. Psychol.*, 55:591–621, 2004.

[52] Nicole Salomons, Sarah Strohkorb Sebo, Meiying Qin, and Brian Scassellati. A minority of one against a majority of robots: robots cause normative and informational conformity. *ACM Transactions on Human-Robot Interaction (THRI)*, 10(2):1–22, 2021.

[53] Christos Kyrlitsias, Despina Michael-Grigoriou, Domna Banakou, and Maria Christofi. Social conformity in immersive virtual environments: The impact of agents' gaze behavior. *Frontiers in psychology*, 11:2254, 2020.

[54] Cäzilia Loibl, Cass R Sunstein, Julius Rauber, and Lucia A Reisch. Which europeans like nudges? approval and controversy in four european countries. *Journal of Consumer Affairs*, 52(3):655–688, 2018.

[55] Tiziana Zalla, Frederique Amsellem, Pauline Chaste, Francesca Ervas, Marion Leboyer, and Maud Champagne-Lavau. Individuals with autism spectrum disorders do not use social stereotypes in irony comprehension. *PloS one*, 9(4):e95568, 2014.

[56] Sergio Melogno, Maria A Pinto, and Margherita Orsolini. Novel metaphors comprehension in a child with high-functioning autism spectrum disorder: a study on assessment and treatment. *Frontiers in Psychology*, 7:2004, 2017.

[57] Salah Hamdoun, Rebecca Monteleone, Terri Bookman, and Katina Michael. Ai-based and digital mental health apps: Balancing need and risk. *IEEE Technology and Society Magazine*, 42(1):25–36, 2023.

[58] Gerhard Stemberger. Psychotherapy: The challenge and power of consistency. *Gestalt Theory*, 43(1):1–12, 2021.

[59] W John Livesley. Moving beyond specialized therapies for borderline personality disorder: the importance of integrated domain-focused treatment. *Psychodynamic Psychiatry*, 40(1):47–74, 2012.